\documentclass[prl,twocolumn,showpacs,floatfix,
nobalancelastpage,
nofootinbib,superscriptaddress]{revtex4}

\usepackage{graphicx}
\usepackage{epstopdf}
\usepackage{color}

\begin{document}

\title{CAST search for sub-eV mass solar axions with $^3$He buffer gas}

\newcommand{\Dogus}{Dogus University, Istanbul, Turkey}
\newcommand{\Saclay}{IRFU, Centre d'\'Etudes Nucl\'eaires de Saclay (CEA-Saclay), Gif-sur-Yvette, France}
\newcommand{\CERN}{European Organization for Nuclear Research (CERN), Gen\`eve, Switzerland}
\newcommand{\INR}{Institute for Nuclear Research (INR), Russian Academy of Sciences, Moscow, Russia}
\newcommand{\MPE}{Max-Planck-Institut f\"{u}r Extraterrestrische Physik, Garching, Germany}
\newcommand{\Trieste}{Instituto Nazionale di Fisica Nucleare (INFN), Sezione di Trieste and Universit\`a di Trieste, Trieste, Italy}
\newcommand{\Zaragoza}{Instituto de F\'{\i}sica Nuclear y Altas Energ\'{\i}as, Universidad de Zaragoza, Zaragoza, Spain }
\newcommand{\Chicago}{Enrico Fermi Institute and KICP, University of Chicago, Chicago, IL, USA}
\newcommand{\Thessaloniki}{Aristotle University of Thessaloniki, Thessaloniki, Greece}
\newcommand{\Demokritos}{National Center for Scientific Research ``Demokritos'', Athens, Greece}
\newcommand{\Freiburg}{Albert-Ludwigs-Universit\"{a}t Freiburg, Freiburg, Germany}
\newcommand{\Patras}{Physics Department, University of Patras, Patras, Greece}
\newcommand{\Athens}{National Technical University of Athens, Athens, Greece}
\newcommand{\MPI}{MPI Halbleiterlabor, M\"{u}nchen, Germany}
\newcommand{\Vancouver}{Department of Physics and Astronomy, University of British Columbia, Vancouver, Canada }
\newcommand{\Darmstadt}{Technische Universit\"{a}t Darmstadt, IKP, Darmstadt, Germany}
\newcommand{\Frankfurt}{Johann Wolfgang Goethe-Universit\"at, Institut f\"ur Angewandte Physik, Frankfurt am Main, Germany}
\newcommand{\Zagreb}{Rudjer Bo\v{s}kovi\'{c} Institute, Zagreb, Croatia}
\newcommand{\MPP}{Max-Planck-Institut f\"{u}r Physik (Werner-Heisenberg-Institut), M\"unchen, Germany}
\newcommand{\LLNL}{Lawrence Livermore National Laboratory, Livermore, CA, USA}
\newcommand{\MPIS}{Max-Planck-Institut f\"{u}r Sonnensystemforschung, Katlenburg-Lindau, Germany}
\newcommand{\Bogazici}{Bogazici University, Istanbul, Turkey}
\newcommand{\Glasgow}{Department of Physics and Astronomy, University of Glasgow, Glasgow, UK}
\newcommand{\PNSensor}{PNSensor GmbH, M\"unchen, Germany}
\newcommand{\XFEL}{European XFEL GmbH, Notkestrasse 85, 22607 Hamburg, Germany}
\newcommand{\ECU}{Excellence Cluster Universe, Technische Universit\"{a}t M\"unchen, Garching, Germany}
\newcommand{\Naval}{Naval Postgraduate School, Monterey, CA, USA}

%\begin{document}

%\title{CAST search for sub-eV mass solar axions extended with $^3$He
%buffer gas}

%\author{Authors}\affiliation{Affiliation}

\affiliation{\Dogus}
\affiliation{\Saclay}
\affiliation{\CERN}
\affiliation{\INR}
\affiliation{\MPE}
\affiliation{\Trieste}
\affiliation{\Zaragoza}
\affiliation{\Chicago}
\affiliation{\Thessaloniki}
\affiliation{\Demokritos}
\affiliation{\Freiburg}
\affiliation{\Patras}
\affiliation{\Athens}
\affiliation{\MPI}
\affiliation{\Vancouver}
\affiliation{\Darmstadt}
\affiliation{\Frankfurt}
\affiliation{\Zagreb}
\affiliation{\MPP}
\affiliation{\LLNL}
\affiliation{\MPIS}

\author{    M.~Arik}\altaffiliation[Present addr.: ]{\Bogazici}\affiliation{\Dogus}
\author{    S.~Aune  }\affiliation{\Saclay}
\author{    K.~Barth  }\affiliation{\CERN}
\author{    A.~Belov  }\affiliation{\INR}
\author{    S.~Borghi  }\altaffiliation[Present addr.: ]{\Glasgow}\affiliation{\CERN}
\author{    H.~Br\"auninger  }\affiliation{\MPE}
\author{    G.~Cantatore  }\affiliation{\Trieste}
\author{    J.~M.~Carmona  }\affiliation{\Zaragoza}
\author{    S.~A.~Cetin  }\affiliation{\Dogus}
\author{    J.~I.~Collar  }\affiliation{\Chicago}
\author{    T.~Dafni  }\affiliation{\Zaragoza}
\author{    M.~Davenport  }\affiliation{\CERN}
\author{    C.~Eleftheriadis  }\affiliation{\Thessaloniki}
\author{    N.~Elias  }\affiliation{\CERN}
\author{    C.~Ezer  }\altaffiliation[Present addr.: ]{\Bogazici}\affiliation{\Dogus}
\author{    G.~Fanourakis  }\affiliation{\Demokritos}
\author{    E.~Ferrer-Ribas  }\affiliation{\Saclay}
\author{    P.~Friedrich    }\affiliation{\MPE}
\author{    J.~Gal\' an  }\altaffiliation[Present addr.: ]{\Saclay}\affiliation{\Zaragoza}
\author{    J.~A.~Garc\' ia  }\affiliation{\Zaragoza}
\author{    A.~Gardikiotis  }\affiliation{\Patras}
\author{    E.~N.~Gazis  }\affiliation{\Athens}
\author{    T.~Geralis  }\affiliation{\Demokritos}
\author{    I.~Giomataris  }\affiliation{\Saclay}
\author{    S.~Gninenko  }\affiliation{\INR}
\author{    H.~G\' omez  }\affiliation{\Zaragoza}
\author{    E.~Gruber  }\affiliation{\Freiburg}
\author{    T.~Guth\"orl  }\affiliation{\Freiburg}
\author{    R.~Hartmann  }\altaffiliation[Present addr.: ]{\PNSensor}\affiliation{\MPI}
\author{    F.~Haug  }\affiliation{\CERN}
\author{    M.~D.~Hasinoff  }\affiliation{\Vancouver}
\author{    D.~H.~H.~Hoffmann  }\affiliation{\Darmstadt}
\author{    F.~J.~Iguaz  }\altaffiliation[Present addr.: ]{\Saclay}\affiliation{\Zaragoza}
\author{    I.~G.~Irastorza  }\affiliation{\Zaragoza}
\author{    J.~Jacoby  }\affiliation{\Frankfurt}
\author{    K.~Jakov\v ci\' c  }\affiliation{\Zagreb}
\author{    M.~Karuza  }\affiliation{\Trieste}
\author{    K.~K\"onigsmann  }\affiliation{\Freiburg}
\author{    R.~Kotthaus  }\affiliation{\MPP}
\author{    M.~Kr\v{c}mar  }\affiliation{\Zagreb}
\author{    M.~Kuster  }\altaffiliation[Present addr.: ]{\XFEL}\affiliation{\MPE}\affiliation{\Darmstadt}
\author{    B.~Laki\'{c}  }\email[Corresponding author: ]{Biljana.Lakic@irb.hr}\affiliation{\Zagreb}
\author{    J.~M.~Laurent  }\affiliation{\CERN}
\author{    A.~Liolios  }\affiliation{\Thessaloniki}
\author{    A.~Ljubi\v{c}i\'{c}  }\affiliation{\Zagreb}
\author{    V.~Lozza  }\affiliation{\Trieste}
\author{    G.~Lutz  }\altaffiliation[Present addr.: ]{\PNSensor}\affiliation{\MPI}
\author{    G.~Luz\'on  }\affiliation{\Zaragoza}
\author{    J.~Morales  }\altaffiliation[Deceased.]{}\affiliation{\Zaragoza}
\author{    T.~Niinikoski  }\altaffiliation[Present addr.: ]{\ECU}\affiliation{\CERN}
\author{    A.~Nordt  }\altaffiliation[Present addr.: ]{\CERN}\affiliation{\MPE}\affiliation{\Darmstadt}
\author{    T.~Papaevangelou  }\affiliation{\Saclay}
\author{    M.~J.~Pivovaroff  }\affiliation{\LLNL}
\author{    G.~Raffelt  }\affiliation{\MPP}
\author{    T.~Rashba  }\affiliation{\MPIS}
\author{    H.~Riege  }\affiliation{\Darmstadt}
\author{    A.~Rodr\' iguez  }\affiliation{\Zaragoza}
\author{    M.~Rosu  }\affiliation{\Darmstadt}
\author{    J.~Ruz  }\affiliation{\Zaragoza}\affiliation{\CERN}
\author{    I.~Savvidis  }\affiliation{\Thessaloniki}
\author{    P.~S.~Silva  }\affiliation{\CERN}
\author{    S.~K.~Solanki  }\affiliation{\MPIS}
\author{    L.~Stewart  }\affiliation{\CERN}
\author{    A.~Tom\' as  }\affiliation{\Zaragoza}
\author{    M.~Tsagri  }\altaffiliation[Present addr.: ]{\CERN}\affiliation{\Patras}
\author{    K.~van~Bibber  }\altaffiliation[Present addr.: ]{\Naval}\affiliation{\LLNL}
\author{    T.~Vafeiadis  }\affiliation{\CERN}\affiliation{\Thessaloniki}\affiliation{\Patras}
\author{    J.~Villar  }\affiliation{\Zaragoza}
\author{    J.~K.~Vogel  }\altaffiliation[Present addr.: ]{\LLNL}\affiliation{\Freiburg}\affiliation{\LLNL}
\author{    S.~C.~Yildiz  }\altaffiliation[Present addr.: ]{\Bogazici}\affiliation{\Dogus}
\author{    K.~Zioutas  }\affiliation{\CERN}\affiliation{\Patras}

\collaboration{CAST Collaboration} \noaffiliation

\date{\today}

\begin{abstract}
The CERN Axion Solar Telescope (CAST) has extended its search for solar axions by
using $^3$He as a buffer gas. At $T=1.8$~K this allows for larger
pressure settings and hence sensitivity to higher axion masses than
our previous measurements with $^4$He. With about 1~h of data taking at each of 252
different pressure settings we have scanned the axion mass range $0.39~{\rm eV}\alt
m_a\alt0.64~{\rm eV}$. From the absence of excess X-rays when the
magnet was pointing to the Sun we set a typical upper limit on the
axion-photon coupling of $g_{a\gamma}\alt\hbox{2.3}\times
10^{-10}~{\rm GeV}^{-1}$ at 95\% CL, the exact value depending on
the pressure setting. KSVZ axions are excluded at the upper end of our mass range,
the first time ever for any solar axion search. In future we will extend our
search to $m_a\alt1.15$~eV, comfortably overlapping with
cosmological hot dark matter bounds.
\end{abstract}


\maketitle

%%%%%%%%%%%%%%%%%%%%%%%%%%%%%%%%%%%%%%%%%%%%%%%%%%%%%%%%%%%%%%%%%%%%%%
%% Introduction %%%%%%%%%%%%%%%%%%%%%%%%%%%%%%%%%%%%%%%%%%%%%%%%%%%%%%
%%%%%%%%%%%%%%%%%%%%%%%%%%%%%%%%%%%%%%%%%%%%%%%%%%%%%%%%%%%%%%%%%%%%%%

{\em Introduction.}---The Peccei-Quinn mechanism is the most
compelling explanation for why in QCD the $\Theta$ term does not
cause measurable CP-violating effects such as a large
neutron electric dipole moment
\cite{Peccei:2006as,Kim:2008hd,Nakamura:2010zzi}. A testable
consequence is the existence of axions, low-mass pseudoscalar bosons
that are closely related to neutral pions. The axion mass is given
by $m_a f_a\sim m_\pi f_\pi$ and the two-photon interaction strength
scales with $f_\pi/f_a$ where $f_\pi\sim92$~MeV is the pion decay
constant and $f_a$ a large energy scale related to the breaking of a
new U(1) symmetry of which the axion is the Nambu-Goldstone boson.

Axions would have been produced in the early universe by the
vacuum realignment mechanism and radiation from cosmic strings,
leading to a cold dark matter component, as well as from thermal interactions,
leading to a hot dark matter
component~\cite{Sikivie:2006ni,Wantz:2009it}. Precision cosmology requires
$m_a\alt0.9$~eV for the latter \cite{Hannestad:2010yi,Cadamuro:2010cz},
with the usual caveats concerning systematic
uncertainties. The cold component increases with decreasing $m_a$
and provides all dark matter for $m_a\sim 10~\mu$eV
($f_a\sim10^{12}$~GeV), with large uncertainties depending on the
early-universe scenario. The ongoing ADMX dark matter
search~\cite{Asztalos:2009yp}, based on Sikivie's idea of
axion-photon conversion in a macroscopic $B$
field~\cite{Sikivie:1983ip}, provides one of the few realistic
opportunities to find ``invisible axions'' \cite{Asztalos:2006kz}.
        
\begin{figure}
\includegraphics[width=1.\columnwidth]{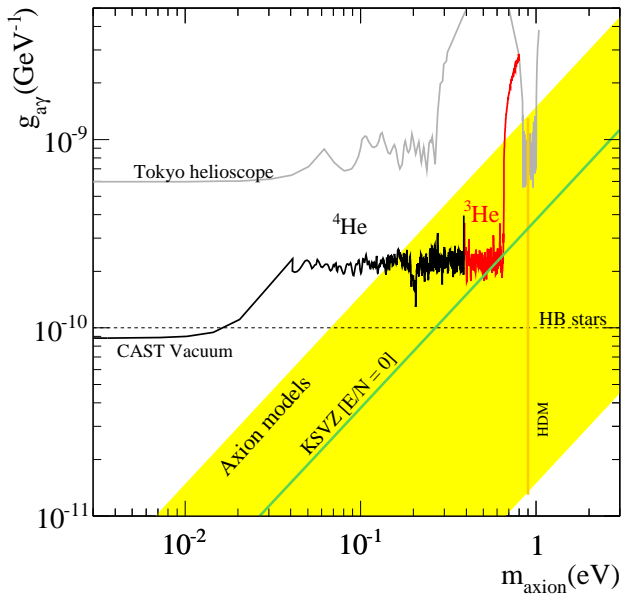}
\caption{Exclusion regions in the $m_a$--$g_{a\gamma}$--plane achieved
by CAST in the vacuum \cite{Zioutas:2004hi,Andriamonje:2007ew},
$^4$He \cite{Arik:2008mq} and $^3$He phase. We also show
constraints from the Tokyo helioscope
\cite{Moriyama:1998kd,Inoue:2002qy,Inoue:2008zp},
horizontal branch (HB)
stars~\cite{Raffelt:2006cw}, and the hot dark matter (HDM)
bound~\cite{Hannestad:2010yi}. The yellow
band represents typical theoretical models with
$\left|E/N-1.95\right|=0.07$--7. The green solid
line corresponds to $E/N=0$ (KSVZ model).}\label{fig:limits}
\end{figure}

Axions would also emerge from the hot interiors of stars, the Sun
being the most powerful ``local'' source~\cite{Raffelt:2006cw}. To
search for these axions, one can
use magnetically induced $a\gamma$ conversion in a dipole magnet
pointing toward the Sun (``axion helioscope'' technique \cite{Sikivie:1983ip}). This
is analogous to neutrino flavor oscillations, $a\gamma$ mixing being
caused by the $B$ field \cite{Raffelt:1987im}.
The axion-photon interaction is given by Lagrangian
\hbox{${\cal L}_{a\gamma}=g_{a\gamma}{\bf E}\cdot{\bf
B}\,a$} with $g_{a\gamma}=(\alpha/2\pi f_a)\,[E/N-2(4+z)/3(1+z)]$.
Here $z=m_u/m_d$ with the canonical value 0.56, although the range
0.35--0.60 is possible~\cite{Nakamura:2010zzi}. $E/N$ is a
model-dependent ratio of small integers \cite{Cheng:1995fd} and
$E/N=0$ (KSVZ model \cite{Kim:1979if,Shifman:1979if}) is our
benchmark case (green line in Fig.~\ref{fig:limits}).

After a pioneering axion helioscope
in Brookhaven~\cite{Lazarus:1992ry}, a fully steerable
instrument was built in
Tokyo~\cite{Moriyama:1998kd,Inoue:2002qy,Inoue:2008zp}. The largest
helioscope yet is the CERN Axion Solar Telescope (CAST), using a
refurbished LHC test magnet ($L=9.26$~m, $B\sim9.0$~T) mounted to
follow the Sun for about 1.5~h both at dawn and dusk
\cite{Zioutas:2004hi,Andriamonje:2007ew,Arik:2008mq,Andriamonje:2009dx,
Andriamonje:2009ar}. CAST began operation in 2003 and after two years of
data taking with vacuum inside the magnet bores achieved a limit of
$g_{a\gamma}<0.88\times10^{-10}~{\rm GeV}^{-1}$ at 95\% CL for
$m_{a} \alt 0.02$~eV \hbox{\cite{Zioutas:2004hi,Andriamonje:2007ew}}.
While these results are excellent to constrain very light axion-like
particles \cite{Jaeckel:2010ni}, realistic QCD axions are not
covered because the $g_{a\gamma}$ bounds quickly degrade for
$m_a\agt0.02$~eV (Fig.~\ref{fig:limits}).

Sensitivity to higher axion masses improves if the conversion
volume contains a buffer gas such as helium~\cite{vanBibber:1988ge}.
Then the $a \gamma$ conversion probability is
\begin{equation} 
P_{a \rightarrow \gamma}    =  \left( \! \frac{Bg_{a \gamma}}{2} \! \right)^{2} 
       \,\, \frac{1 \! + \! e^{-\Gamma L} \! - \! 2 e^{-\Gamma L/2}
 \cos(qL)}{q^{2} \! + \! \Gamma^{2} \! /4} 
        \label{prob}
\end{equation}
where $\Gamma$ is the inverse photon absorption length in the buffer gas, while
the momentum difference between the $a$ and $\gamma$ propagation
eigenstates is given by
$q^2=[(m_a^2-m_\gamma^2)/2E]^2+(g_{a\gamma}B)^2$. For
$m_a^2=m_\gamma^2$, axions and photons are maximally mixed and reach
$P_{a\to\gamma} = (g_{a\gamma}B L/2)^2=1.7\times10^{-17}$ for $L=9.26$~m,
$B=9.0$~T and $g_{a\gamma}=10^{-10}~{\rm GeV}^{-1}$.
For $m_a\not=m_\gamma$, the conversion probability rapidly decreases
due to the axion-photon momentum mismatch.

The maximum  $P_{a\to\gamma}$ can be restored by matching $m_a$ with
a photon refractive mass $m_\gamma$ \cite{vanBibber:1988ge}. This
method was first applied by the Brookhaven helioscope using $^4$He as a
buffer gas \cite{Lazarus:1992ry} and later allowed CAST to reach
realistic axion models for $m_a\alt0.4$~eV (Fig.~\ref{fig:limits})
\cite{Arik:2008mq}. However, $T=1.8$~K of the superconducting magnet
restricts, due to condensation, the maximum $^4$He pressure to $\sim 14$~mbar  
thus allowing us to scan axion masses $m_{a}\alt0.4$~eV. To close
the gap to the hot-dark matter bound, we have used $^3$He as buffer
gas to allow CAST to search up to $m_{a}\alt1.15$~eV. The first results
from this novel technique for the axion mass range $0.39 \alt m_{a} \alt 0.64$~eV
are reported here.

{\em Upgrades.}---
After completing the data taking with $^4$He as a buffer gas,
the CAST experiment performed several upgrades in order to prepare for
data taking with $^3$He. The most important upgrade was the design and
installation of a sophisticated $^3$He gas system.

To scan over a range of axion masses, CAST needs to control precisely the helium gas density in the cold bores. 
This is achieved by filling the cold bores with precisely metered amount of gas in incremental steps.
The step size of the gas density is equivalent to a pressure change of between 
$\rm{0.083\,mbar}$ and $\rm{0.140\,mbar}$ (calculated for gas at nominal temperature of
$\rm{1.8\,K}$). To scan the whole available mass range efficiently, data taking runs
cover two density settings per solar tracking. During the measurement, 
it is desired to have the gas density in the cold bores to be as homogenous and as stable as possible.
The density homogeneity is ensured by the excellent thermal coupling with the superfluid
helium bath surrounding the cold bores.
To achieve the proper densities in the cold bores of CAST, and to be able to reproducibly refill
the bores (allowing us to search the same axion mass), requires that the gas system is capable of
adjusting to fluctuations of external conditions (e.g. variations of the room and magnet temperatures).
The density stability due to uncorrelated temperature fluctuations
is met by minimizing the volume of external pipework connected to the cold bore.
The density fluctuations are well within the density stability limit of $0.001 \,\, \rm{ kg}/\rm{m}^{3}$ 
(for example, the allowed magnet temperature fluctuations are about 350~mK while typical fluctuation
during magnet vertical movement is 35~mK).

The $^3$He system can be described as a hermetically closed gas circuit which is
divided into functional sections with specific purposes:
Storage,
Trap purge system,
Metering and ramping of gas density,
Expansion volume,
Recovery and circulation.
        
All the necessary helium for CAST physics runs is transferred 
to the storage volume that has been specifically engineered to
keep the gas pressure below atmospheric. Before entering
the metering volumes, the gas passes through two charcoal traps.
The first one at ambient temperature traps oil and water vapour
while the second at liquid nitrogen temperature removes residual gases. 

The metering precision of the gas density is obtained by the accurate temperature control
of the metering volumes, and by use of a metrology-grade pressure-measuring instruments
to determine the amount of gas introduced into the cold bores. This amount of gas is calculated
by accurately measuring the pressure decrease in the metering volumes. 
The reproducibility for the amount of gas sent from the metering volume into the magnet is $\rm{61\,ppm}$.

The gas is confined in the cold bore region of the magnet with thin X-ray windows installed on both
ends.
The windows are made of 15~$\mu$m-thick polypropylene stretched over a mostly-open strongback
structure to provide high X-ray transmission, resistance to a sudden rise in pressure and minimal
helium leakage.  Heaters on the window flanges allow for periodic bake-out of gases adsorbed on the polypropylene.

In case of quench, a sudden loss of superconductivity in the magnet,
the temperature of the magnet increases rapidly.
If the cold volume remains closed, the gas pressure
abruptly increases and endangers the integrity of the X-ray windows.
The windows can safely withstand pressures up to 1.2~bar, and to prevent
rupture during a quench, the system must safely evacuate the $^3$He from the
cold bores to the expansion volume.  Thus, the expansion volume, initially
under vacuum, acts as a buffer reservoir for the gas that is intentionally
expelled from the cold bores.
The CAST $^3$He system will be described in detail in a future publication.
         
It is a demanding task to compute the amount of gas needed to achieve the desired gas density.
In fact, such calculations can only reliably be performed through computational fluid dynamic (CFD)
simulations that account for the as-built system, as well as different physical phenomena
such as hydrostatic effects, convection and buoyancy.
For a typical run, e.g. $m_{\gamma}=0.64$~eV, the intrinsic mass-acceptance width 
coming from the coherence condition \cite{Arik:2008mq} increases due to the mentioned phenomena
from 0.8~meV  to 1.6~meV while the height decreases accordingly.
The CFD simulations will be described in detail in a future publication.     
        
During preparations for the $^3$He data taking, the CAST X-ray detectors were upgraded as well.
The Time Projection Chamber (TPC) with a multi-wire proportional readout \cite{Autiero:2007uf} 
that had covered both bores of the sunset end of the magnet was replaced by two Micromegas
detectors of similar dimensions of the one previously installed at the sunrise side \cite{Abbon:2007ug}
but with readouts fabricated with novel bulk and microbulk techniques 
\cite{Andriamonje:2010zz, Galan:2010zz, Aune:2009zzc}. On the sunrise end a new shielded bulk
(and later on microbulk) Micromegas replaced the unshielded one of our previous run \cite{Abbon:2007ug}.
These novel techniques provide several improvements in terms of stability and homogeneity of response,
energy resolution, simplicity of construction \cite{Andriamonje:2010zz, Galan:2010zz, Aune:2009zzc}
and, for the case of microbulk readouts, material radiopurity \cite{Cebrian:2010ta}. This is the first
time these kinds of readouts are used in a physics run of a low background experiment.
These new Micromegas detectors have obtained background levels down to $\sim5\times 10^{-6}$ 
counts~keV$^{-1}$cm$^{-2}$s$^{-1}$ in the energy range of interest, one order of magnitude better that their
predecessors \cite{Arik:2008mq}. This improvement is due to new shielding in the case of the sunrise
detector, and to better rejection capabilities of the Micromegas readout with respect to the MWPC one,
for the sunset set-up.
The remaining background is attributed to unshielded external gammas (mostly due to
the solid angle of incomplete shielding on the side where the detector is connected
to the magnet bore).
The X-ray mirror telescope with a pn-CCD chip \cite{Kuster:2007ue} covering the
other bore of the sunrise side remained unchanged.

{\em Data analysis and results.}---
Data presented in this paper correspond to the first 252 density
steps of the $^3$He phase, which encompass an equivalent
axion mass range between 0.39 eV and 0.64 eV. The total available exposure
time in axion-sensitive conditions is about 200
hours per detector, shared approximately equally among each of
the four CAST detectors, as well as among the stated range of axion masses.

Data analysis is performed in a manner similar to our previous results
obtained with $^4$He gas.  This time, however, we use an unbinned likelihood
function that can be expressed as
\begin{equation}\label{unbinned}
\log L \propto -R_T + \sum_i^N \log R(t_i, E_i, d_i)
\end{equation}
\noindent where the sum runs over each of the $N$ detected counts
and $R(t_i, E_i, d_i)$ is the event rate expected at the time $t_i$, energy
$E_i$ and detector $d_i$ of the event $i$. $R_T$ is the
integrated expected number of counts over all exposure time,
energy and detectors
\begin{equation}\label{R}
R(t, E, d) = B_d + S(t,E,d)
\end{equation}
\noindent where $B_d$ is the background rate of detector $d$.
$S(t,E,d)$ is the expected rate from axions in detector $d$ which
depends on the axion properties $g_{a\gamma}$ and $m_a$
\begin{equation}\label{S}
S(t,E,d) = \frac{d\Phi_a}{dE} P_{a\rightarrow \gamma} \epsilon_d
\end{equation}
\noindent where  $P_{a\rightarrow \gamma}$ is the axion photon conversion probability
in the CAST magnet~(\ref{prob}), $\epsilon_d$ the detector efficiency, and
\begin{equation}
 \frac{d\Phi_a}{dE}  =  6.02 \times 10^{10} \,  g_{10}^{2} \,
  \frac{E^{2.481}}{e^{E/1.205}} \, \, \, \, \, \,  \mathrm{cm}^{-2} \, \mathrm{s}^{-1} \, \mathrm{keV}^{-1} 
\label{spectrum}
\end{equation}
is the solar axion spectrum, with $g_{10}=g_{a \gamma}/(10^{-10} \, \mathrm{GeV}^{-1})$ and energies in keV.

As explained in \cite{Arik:2008mq}, the $m_a$ dependency of the above expression is encoded in the probability $P_{a\rightarrow \gamma}$, which is coherently enhanced for values of $m_a$ matching the photon mass $m_\gamma$ induced by the buffer gas density, while it is negligible for values away from $m_\gamma$. Therefore, only the counts observed with the gas density matching a given axion mass $m_a$ will contribute to the $\log L$ (and the exclusion plot) for that mass $m_a$.

The use of the unbinned likelihood (\ref{unbinned}), instead of the binned one used in our previous result \cite{Arik:2008mq} is
motivated by the overall reduction of background rates achieved by CAST detectors with respect to the ones of the $^4$He phase, as well as due to the reduced $^3$He density setting exposure time
(one half that for $^{4}$He)       
due to time constraints of the overall data taking campaign. Indeed, the effective number of background counts in this analysis is about 1 count per density step for the Micromegas detectors, and about 0.2 in the fiducial spot of the CCD/Telescope system.
Because of that, the result obtained is almost statistics limited, and further background
reduction would give only slightly better sensitivity unless longer exposure times are available.
        
\begin{figure}
\includegraphics[width=1.\columnwidth]{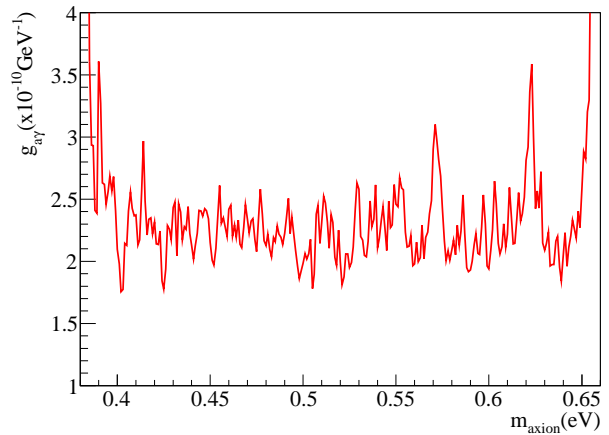}
\caption{Expanded view of the limit achieved in the $^3$He CAST phase for
        axion mass range between 0.39 eV and 0.64 eV.}\label{fig:zoom}
\end{figure}

The remaining process is similar to the one followed in our previous
results \cite{Arik:2008mq}: a best fit value $g_{\mathrm{min}}^4$ is obtained after
maximization of $L$ (for a fixed value of $m_a$). The obtained value is compatible with
the absence of positive signal, and therefore an upper
limit $g_{95}^4$ is obtained by integration of the Bayesian probability from
zero up to 95\% of its area in $g^4$. This value is computed for
many values of the axion mass $m_a$ in order to configure the full
exclusion plot shown in Fig.~\ref{fig:limits}. A close up of the same exclusion
plot is shown in Fig.~\ref{fig:zoom}, focused specifically in the axion mass range
which has been explored in the data presented here.

As can be seen in Fig.~\ref{fig:limits}, CAST extends its previous exclusion plot
towards higher axion masses, excluding the interval 0.39--0.64 eV down
to an average value of the axion-photon coupling of
$2.27\times 10^{-10}$ GeV$^{-1}$. The actual
limit contour has high-frequency structure that is a result of statistical fluctuations
that occur when a limit is computed for a specific mass using only a few hours of data.      
        
%\clearpage

{\it Conclusions.}---CAST has taken a great leap forward by using
$^3$He as buffer gas to cover $m_a$ in the gap between our $^4$He
results and the hot dark matter bound.
It is the first axion helioscope ever that has
crossed the ``axion line'' for the benchmark KSVZ case. After covering $0.39~{\rm
eV}\alt m_a\alt0.64~{\rm eV}$ we will eventually reach 1.15~eV with the
$^3$He setup. If axions are not detected by CAST, the next challenge is to
move down in the $m_a$--$g_{a\gamma}$ plot below the ``axion band''
of theoretical models. Such a goal cannot be achieved with the
existing CAST apparatus and will require significant improvements of
detector and magnet properties~\cite{Irastorza:2011gs,Baker:2011na}
 or a completely new approach.

{\it Acknowledgments.}---We thank CERN for hosting the experiment
and for the technical support to operate the magnet and cryogenics.
We thank the CERN CFD team for their essential contribution to the CFD work.
We acknowledge support from NSERC (Canada), MSES (Croatia) under the grant number
098-0982887-2872, CEA (France), BMBF (Germany) under the grant numbers 05 CC2EEA/9
and 05 CC1RD1/0 and DFG (Germany) under grant numbers HO 1400/7-1 and EXC-153,
the Virtuelles Institut f\"ur Dunkle Materie und Neutrinos -- VIDMAN (Germany),
GSRT (Greece), RFFR (Russia), the Spanish Ministry of Science and Innovation (MICINN)
under grants FPA2007-62833 and FPA2008-03456, Turkish Atomic Energy Authority
(TAEK), NSF (USA) under Award number 0239812,
US Department of Energy, NASA under the grant number NAG5-10842. Part of this work
was performed under the auspices of the US Department of Energy by 
Lawrence Livermore National Laboratory under Contract DE-AC52-07NA27344. 
We acknowledge the helpful discussions within the network on direct dark matter
detection of the ILIAS integrating activity (Contract number: RII3-CT-2003-506222).
        
%%%%%%%%%%%%%%%%%%%%%%%%%%%%%%%%%%%%%%%%%%%%%%%%%%%%%%%%%%%%%%%%%%%%%%%

%%%%%%%%%%%%%%%%%%%%%%%%%%%%%%%%%%%%%%%%%%%%%%%%%%%%%%%%%%%%%%%%%%%%%%%
\end{document}